\documentclass[aps,prl,twocolumn,amsmath,amssymb,showpacs, superscriptaddress,notitlepage,longbibliography,10pt]{revtex4-1}

\usepackage{graphicx}
\usepackage{physics}
\usepackage{braket}
\usepackage{times}
\usepackage{multirow}
\usepackage{ascmac}
\usepackage{amsthm}
\usepackage{float}
\usepackage{color}
\usepackage{comment}
\usepackage{mathtools}  
\usepackage{mathrsfs} 
\usepackage[colorlinks=True,urlcolor=blue,linkcolor=blue,citecolor=blue]{hyperref}
\usepackage{xcolor}
\usepackage{multirow}
\usepackage{setspace}

\usepackage[toc,page]{appendix}

\usepackage{amsfonts}
\usepackage{amsmath}
\usepackage{amssymb}
\usepackage{bm}

\usepackage{slashed}
\usepackage{dsfont}

\begin{document}
\title{Non-Hermiticity Induced Universal Anomalies in Kondo Conductance}
\author{Wei-Zhu Yi}
\affiliation{National Laboratory of Solid State Microstructures and Department of Physics, Nanjing University, Nanjing 210093, China}
\affiliation{State Key Laboratory of Quantum Functional Materials and Department of Physics, Southern University of Science and Technology, Shenzhen 518055,  China}
\author{Yun Chen}
\affiliation{National Laboratory of Solid State Microstructures and Department of Physics, Nanjing University, Nanjing 210093, China}
\author{Jun-Jun Pang}
\affiliation{National Laboratory of Solid State Microstructures and Department of Physics, Nanjing University, Nanjing 210093, China}
\author{Hong Chen}
\affiliation{National Laboratory of Solid State Microstructures and Department of Physics, Nanjing University, Nanjing 210093, China}
\author{Baigeng Wang}
\email{bgwang@nju.edu.cn}
\affiliation{National Laboratory of Solid State Microstructures and Department of Physics, Nanjing University, Nanjing 210093, China}
\affiliation{Collaborative Innovation Center of Advanced Microstructures, Nanjing University, Nanjing 210093, China}
\affiliation{Jiangsu Physical Science Research Center, Nanjing University, Nanjing 210093, China}
\author{Rui Wang}
\email{rwang89@nju.edu.cn}
\affiliation{National Laboratory of Solid State Microstructures and Department of Physics, Nanjing University, Nanjing 210093, China}
\affiliation{Collaborative Innovation Center of Advanced Microstructures, Nanjing University, Nanjing 210093, China}
\affiliation{Jiangsu Physical Science Research Center, Nanjing University, Nanjing 210093, China}
\affiliation{Hefei National Laboratory, Hefei 230088, People's Republic of China }

\begin{abstract}
Strong correlation, when combined with dissipation in open systems, can lead to a variety of exotic quantum phenomena.  Here, we study nontrivial interplays between non-Fermi liquid behaviors emerging from strong correlation and non-Hermiticity arising from open systems. We propose a practical physical setup that realizes a non-Hermitian multichannel Kondo model. We identify a weak-coupling local moment fixed point and a strong-coupling non-Fermi liquid fixed point under \(\mathcal{PT}\) symmetry, both are enriched by the non-Hermitian effect. Remarkably, universal unconventional Kondo conductance behaviors are found for both cases, which are distinct from all previously studied Kondo systems. Particularly, we show that an anomalous upturn of conductance could take place with increasing the temperature, originating from the interplay between non-Fermi liquid and non-Hermiticity. Our results identify a novel class of transport phenomena unrecognized before, driven by intertwined effects of correlation and dissipation.
\end{abstract}
\maketitle

{\color{blue}\emph{Introduction.}}--Non-Hermitian physics has recently provided fresh insights into condensed matter systems. It leads to a wealth of novel quantum phenomena, including the non-Hermitian skin effect \cite{zhao_two-dimensional_2025,yoshida_non-hermitian_2024,song_non-hermitian_2019-1,okuma_topological_2020,kawabata_entanglement_2023,li_critical_2020,PhysRevLett.124.250402,manna2023,PhysRevLett.124.056802}, non-Hermitian topology \cite{ochkan_non-hermitian_2024,yao_edge_2018,bergholtz_exceptional_2021,yang_homotopy_2024,song_non-hermitian_2019,hamanaka_non-hermitian_2024,kawabata_symmetry_2019,gong_topological_2018}, and exceptional points \cite{lee_exceptional_2022,ding_non-hermitian_2022,tang_exceptional_2020,miri_exceptional_2019}. These phenomena predominantly arise in non-interacting or weakly interacting systems, where quasi-particle descriptions are well-defined and applicable. However, for strongly correlated systems where the notion of quasi-particle fundamentally breaks down, the role of non-Hermiticity remains largely unexplored. 

A typical class of correlated phenomena without quasi-particle descriptions is the non-Fermi liquid (nFL). It is characterized by anomalous transport properties that deviate from conventional Fermi liquids \cite{aj_non-fermi_1999,coleman_how_2001,varma_singular_2002}, which have been observed in various systems, including heavy fermion materials \cite{andres_4f-virtual-bound-state_1975,si_heavy_2010,stewart_non-fermi-liquid_2001,lohneysen_non-fermi-liquid_1994,coleman_theories_1999,custers_break-up_2003,PhysRevLett.127.026401,PhysRevLett.114.177202,PhysRevLett.134.116605,chang_mobius_2017}, high-$T_c$ superconductors \cite{varma_phenomenology_1989,lee_doping_2006,zhang_high-temperature_2024,jiang_interplay_2023,yuan_scaling_2022,cao_strange_2020,chudnovskiy_superconductor-insulator_2022}, etc. Among them, the multichannel Kondo effect has attracted enormous interest, as it serves as an experimentally feasible platform for realizing exotic nFL behaviors \cite{iftikhar_tunable_2018,iftikhar_two-channel_2015,potok_observation_2007,keller_universal_2015,seaman_evidence_1991}.  Notably, the Majorana-based setup proposed by B\'eri and Cooper realizes the multichannel Kondo model with exact channel symmetry without fine-tuning \cite{beri_topological_2012}, resulting in a nFL fixed point (FP) exhibiting fractionalized impurity entropy \cite{PhysRevLett.128.146803,PhysRevLett.129.227703,PhysRevLett.130.146201}. Being fractionalized in nature, this nFL fixed point generates anomalous temperature dependence of thermodynamics, in sharp contrast with the Fermi liquid phase. It is intriguing to ask whether it could drive new quantum phenomena when non-Hermiticity (nH) comes into play? This motivates an unexplored area--the interplay between nFL and nH. 

\begin{figure}
\includegraphics[width=\linewidth]{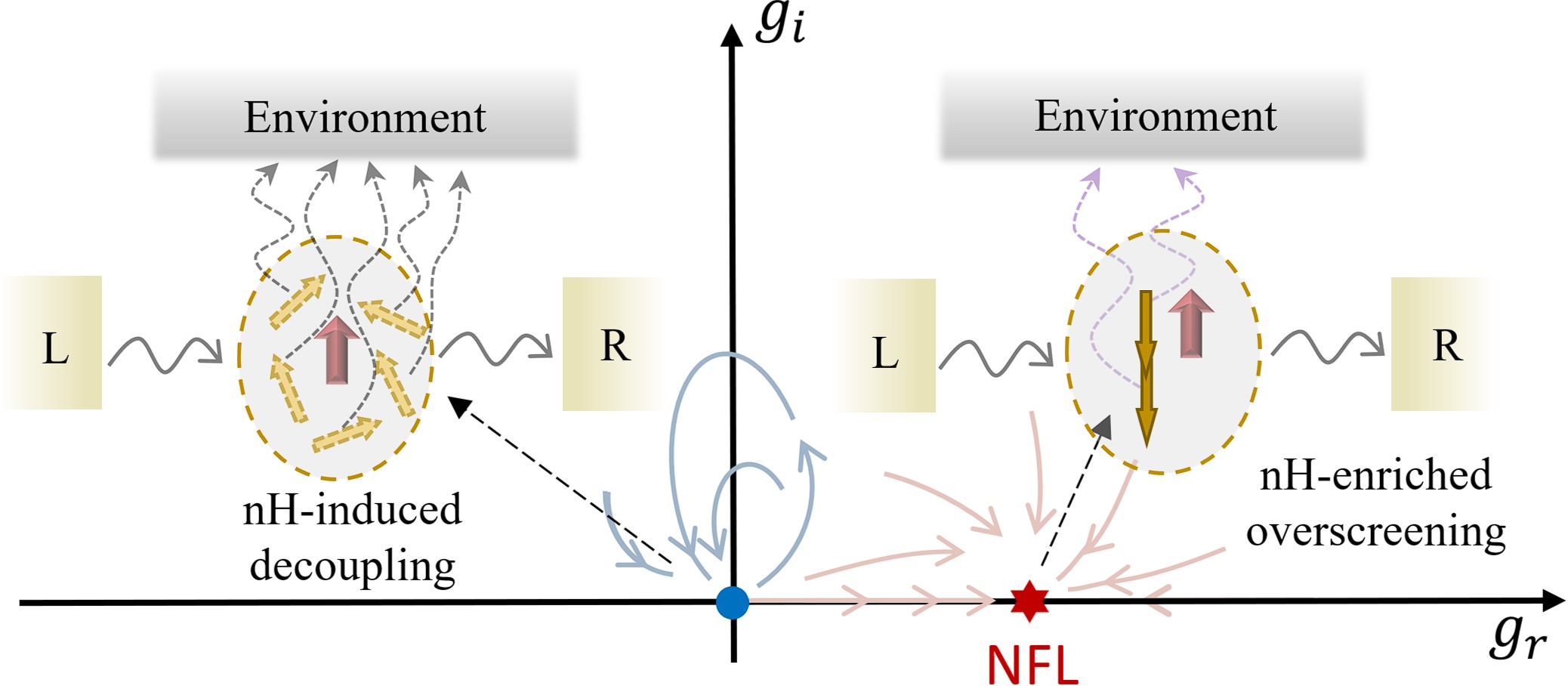}
\caption{\label{fig1} Schematic plot of the two FPs of the nHMCK model revealed by the renormalization group flow. The blue and red dot represent the weak- and strong-coupling FP, respectively. Around both FPs, novel Kondo conductance emerges due to the non-Hermitian effect arising from the dissipation into the environment.}
\end{figure}
 
In this letter, we uncover that the interplay between nFL and nH can generate anomalous Kondo conductance behaviors absent in all existing Kondo systems. To ground our theory in physical settings, we propose a practical setup that realizes the non-Hermitian multichannel Kondo (nHMCK) model (Eq.\eqref{eq3} below). The perturbative renormalization group (RG) analysis predicts both a weak-coupling and a strong-coupling phase, as shown in Fig.\ref{fig1}. Around the weak-coupling FP, the impurity moment decouples from the conduction electrons even for antiferromagnetic Kondo couplings, leading to a universal low-temperature conductance, $\sim1/\mathrm{ln}^2(T/T_K)$. Although the conductance is similar to that of the underscreened Kondo models  ~\cite{posazhennikova_anomalous_2005,coleman_singular_2003,mehta_regular_2005}, it is driven by a completely different dissipation mechanism--the nH-induced decoupling. Beyond the weak-coupling phase, Bethe ansatz approach gives richer physics than that predicted by the perturbative RG analysis \cite{PhysRevB.111.L201106,PhysRevB.111.224407}, which is verified by our non-Hermitian numerical renormalization group (NRG) approach. Importantly, after imposing \(\mathcal{PT}\) symmetry, it is found that the low-energy physics of the strong coupling phase behaves similarly with that of the Hermitian case,  and is described by a boundary conformal field theory (BCFT). Remarkably, the fractionalized nature of the nFL, when combined with non-Hermiticity, causes an anomalous deviation in the Kondo conductance with raising the temperature (Fig.\ref{fig4} below).  Such a nH-enriched anomaly stems from an emergent BCFT with non-Hermitian boundary operators, which is unique to systems exhibiting both nH and nFL. These results point to a novel class of transport phenomena unrecognized before, driven simultaneously by correlation and dissipation.

{\color{blue}\emph{Non-Hermitian Majorana tunneling through dissipative quantum dots.}}--The prototypical Hamiltonian that hosts both nFL and nH physics  is the nHMCK model, i.e., 
 \begin{equation}\label{eq3}
H=H_{\mathrm{lead}}+
\sum^n_{k=1}\sum_{l=1,2,3}J^{\prime}_ls_lc^{\dagger}_{k,\alpha}\tau_{l,\alpha\beta}c_{k,\beta},
\end{equation}
where $s_l$ is the impurity spin operator, $\tau_{l}$ is the 2 by 2 Pauli matrices, $c_{k,\alpha}$ denotes the conduction electrons of ``spin" $\alpha$ and channel $k$, $H_{\mathrm{lead}}$ describes the conduction electrons, $J^{\prime}_l$ is the complex Kondo coupling along the $l$-direction and $J^{\prime}_l=J^{\prime}$ for the isotropic case. In the following, we will  present a two-step realization of this model based on Majorana tunneling junctions. 
%The first step is to introduce non-Hermiticity into the tunneling of Majorana fermions \cite{PhysRevA.93.022102}, and the second is to achieve the Kondo effect via the multi-junction structure \cite{beri_topological_2012}.

\begin{figure}
\includegraphics[width=\linewidth]{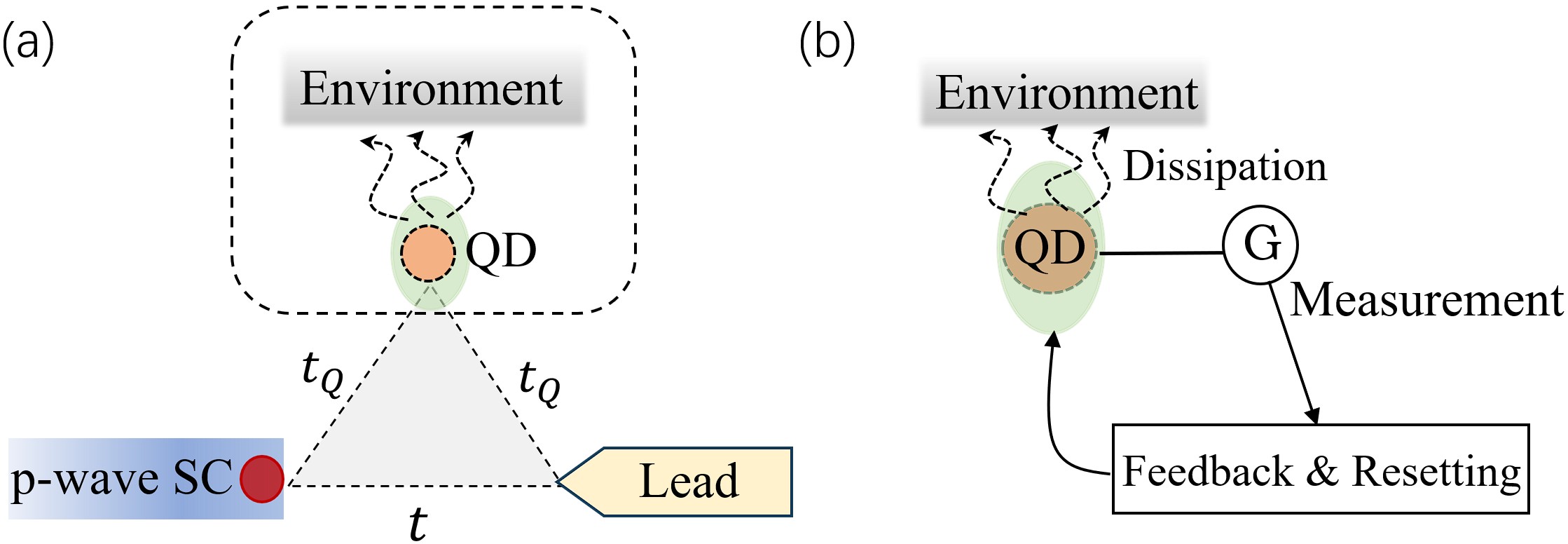}
\caption{\label{fig2} The junction setup that realizes the non-Hermitian Majorana tunneling. (a) Majorana edge mode (red dot) tunnels into the normal lead with the assistance of a QD, which is in turn coupled to an environment. (b) The detailed structure of the QD-environment system, which is attached to a feedback-resetting setup.}
\end{figure}

In the first step, we consider the tunneling junction shown by Fig.\ref{fig2}(a), where the Majorana edge mode tunnels into a normal lead with the assistance of a dissipative quantum dot (QD). In contrast to the QD-induced Kondo effect \cite{PhysRevLett.87.176601,PhysRevLett.101.246805,PhysRevB.66.155308}, here the QD is introduced to involve the dissipation effect by coupling to environment modes~\cite{PhysRevLett.130.200404,PhysRevA.100.053605,PhysRevLett.123.193605,daley_quantum_2014} (Fig.\ref{fig2}(a)). The setup is further attached to a monitoring-feedback device, which resets the particle loss events and ensures the particle conservation of the QD (Fig.\ref{fig2}(b)). 
The QD, described by $H=\sum_i\varepsilon_id^{\dagger}_id_i$ with $\varepsilon_i$ being the $i$-th QD energy level , has a single-body loss into the environment.  This results in a non-unitary dynamics, where the evolution of the QD density matrix $\varrho$ is  governed by the Lindblad master equation,
\begin{equation}\label{dynamics}
    \frac{\mathrm{d}\varrho(t)}{\mathrm{d}t}=-i\left[H^{\prime},\varrho(t)\right]+\sum_{i}\left[L_{i}\varrho(t)L_{i}^{\dagger}-\frac{1}{2}\left\{ L_{i}^{\dagger}L_{i},\varrho(t)\right\} \right],
\end{equation}
where $H^{\prime}$ is the Lamb-shifted QD Hamiltonian renormalized by the environment.  The jump term describes the single-body leakage into the environment~\cite{PhysRevLett.130.200404,PhysRevA.100.053605,PhysRevLett.123.193605}, i.e., $L_i=\sqrt{\Gamma}d_i$ with $\Gamma$ being the decay rate. The particle leakage event can be monitored by an attached empty reservoir~\cite{PhysRevB.111.125157,daley_quantum_2014}. As long as any particle is detected, the QD is immediately reset by the feedback equipment. This ultimately results in a postselection on the Lindbladian,  giving rise to an effective non-Hermitian QD Hamiltonian ~\cite{PhysRevB.109.235139,PhysRevLett.124.196401}, i.e.,

\begin{equation}\label{eqdissQD}
H_{\mathrm{QD,eff}}=\sum_i(\tilde{\varepsilon}_d-i\frac{\Gamma}{2})d^{\dagger}_id_i.
\end{equation}
It is clear that an imaginary chemical potential is generated, realizing a dissipative QD. 

Then, we consider the tunneling of Majorana fermions via the dissipative QD, as shown by Fig.\ref{fig2}(a).  The junction Hamiltonian reads as, $H_{\mathrm{junc}}= H_{\mathrm{QD,eff}}+H_{\mathrm{s-QD}}+ H_{\mathrm{s}}$. It consists of the tunneling processes involving the QD, $H_{\mathrm{s-QD}}=\sum_it_{Q}(\gamma+\psi^{\dagger})d_i+h.c.$, as well as the Majorana-lead tunneling, $H_{\mathrm{s}}=(t\gamma\psi^{\dagger}+h.c.)+H_{\mathrm{lead}}$, where $\gamma$ denotes the Majorana mode,  $\psi$ represents the conduction electron at the tunneling point, $t_Q$ and $t$ are the Majorana and QD tunneling amplitudes, respectively. 

The Schrödinger equation $H_{\mathrm{junc}}|\Psi\rangle=E|\Psi\rangle$ can be written into a coupled equation in the direct product space of the QD and the rest. Tracing out the QD degrees of freedom then generates an effective junction Hamiltonian renormalized by the dissipative QD, $H_{\mathrm{junc,eff}}= H_{\mathrm{s}}+H_{\mathrm{s-QD}}(E-H_{\mathrm{QD,eff}})^{-1}H^{\mathrm{T}}_{\mathrm{s-QD}}$ . After insertion of Eq.\eqref{eqdissQD}, we find that the original tunneling amplitude is renormalized in a nontrivial way. The tunneling from the Majorana to the lead ($t$) and the reversed tunneling $(t^{\star})$ are respectively modified as ,  
\begin{equation}
t^{\prime}_+=t+\frac{|t_{\mathrm{Q}}|^2}{E-\tilde{\varepsilon}_d+i\Gamma/2},~~~~ t^{\prime}_-=t^{\star}+\frac{|t_{\mathrm{Q}}|^2}{E-\tilde{\varepsilon}_d+i\Gamma/2}.
\end{equation}
Clearly,  the tunneling amplitude becomes non-Hermitian because  $t^{\prime\star}_+\neq t^{\prime}_-$. In the gauge where $t$ is real, both $t^{\prime}_{+}$ and $t^{\prime}_{-}$ are reduced to $t^{\prime}=t_r^{\prime}+it_i^{\prime}$, and $t_r^{\prime}=t-\frac{|t_Q|^2\tilde{\varepsilon}_d}{\tilde{\varepsilon}^2_d+(\Gamma/2)^2}$,  $t_i^{\prime}=\frac{|t_Q|^2\Gamma/2}{\tilde{\varepsilon}^2_d+(\Gamma/2)^2}$ \footnote{The QD is assumed to have a higher energy scale than the junction system, i.e. $\tilde{\varepsilon}_d\gg E$}.  Thus, the dissipative QD endows the tunneling process of Majorana fermions with non-Hermiticity.

\begin{figure}
\includegraphics[width=\linewidth]{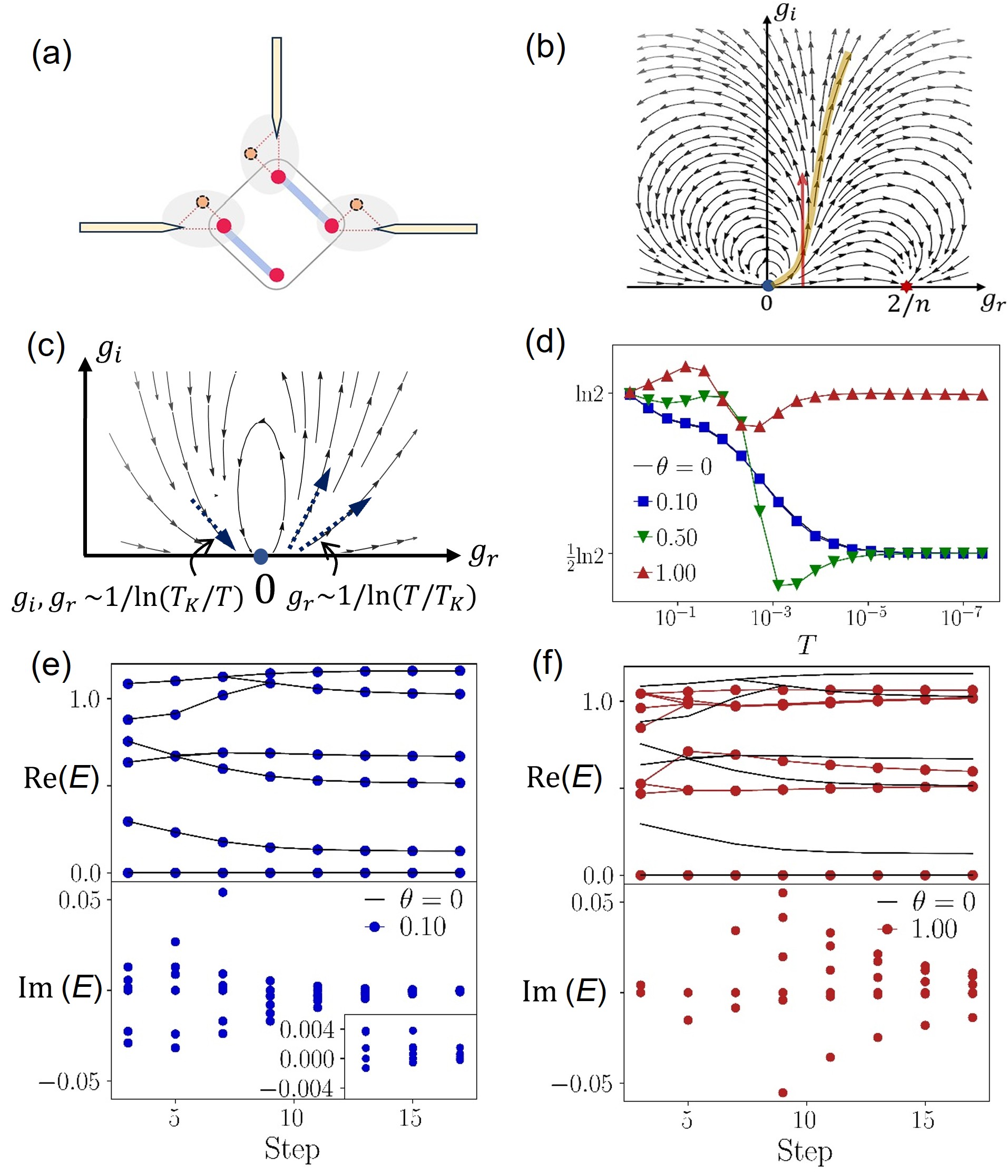}
\caption{\label{fig3} (a) The multi-junction structure that realizes the nHMCK model. The $M=3$ case is shown as an example. (b) The RG flow in the complex plane. The yellow curve denotes the critical points separating the weak- and strong-coupling phase. (c) shows the zoom-in RG flow around the weak-coupling FP, which indicates universal scalings of the Kondo coupling $g_r$ and $g_i$. (d)-(f) The NRG results of the general non-Hermitian Kondo model for $n=2$. (d) shows the impurity entropy versus temperature with varying  $\theta$, and (e), (f) are the energy spectrum for the weak ($\theta=0.1$) and strong non-Hermiticity ($\theta=1.0$) case, respectively.  Both the real and imaginary spectrum are plotted, which are compared to the Hermitian case (black curves).}
\end{figure}

{\color{blue}\emph{Non-Hermitian multichannel Kondo model.}}-- In the second step, we realize the nHMCK model via a multi-junction structure \cite{beri_topological_2012}. As shown by Fig.\ref{fig3}(a), we consider a mesoscopic superconducting island that supports $M_{\mathrm{tot}}$ Majorana modes, i.e., $\gamma_{\alpha}$ \cite{beri_topological_2012,beri_majorana-klein_2013}. Among them, $M$ Majorana modes are coupled to normal leads via the dissipative QD, exhibiting complex tunneling amplitude $t_{\alpha}$. The island is connected to the ground by a capacitor, which contributes to a charging energy $H_c(N)=E_c(N-\frac{q}{e})^2$, where $N$ is the number of electrons on the island, and $q$ is the background charge determined by the voltage across the capacitor.

For $E_c\gg|t_{\alpha}|$, the Schrieffer-Wolff transformation leads to the non-Hermitian effective Hamiltonian,
\begin{equation}\label{eq2}
H=H_{\mathrm{lead}}+\frac{1}{2}\sum_{\alpha\neq\beta}\lambda_{\alpha\beta}\gamma_{\alpha}\gamma_{\beta}\psi^{\dagger}_{\beta}\psi_{\alpha}=H_{\mathrm{lead}}+\sum_{l} J_{l}s_{l}\cdot S_{l},
\end{equation}
where $\psi_{\alpha}$ denotes the electron operator of the $\alpha$-th lead at the tunneling point, $s_{l}=-i\epsilon_{l\alpha\beta}\gamma_{\alpha}\gamma_{\beta}$ and $S_{l}=-i\epsilon_{l\alpha\beta}\psi^{\dagger}_{\beta}\psi_{\alpha}$, with $\epsilon_{l\alpha\beta}$ being the antisymmetric Levi-Civita tensor.  The effective couplings, $\lambda_{\alpha\beta}=[\frac{1}{H_c(N+1)-H_c(N)}+\frac{1}{H_c(N-1)-H_c(N)}]t_{\alpha}t_{\beta}$, $J_{\lambda}=|\epsilon_{\lambda\alpha\beta}|\lambda_{\alpha\beta}$, are both complex, and $\lambda_{\alpha\beta}$ can be written into $\lambda_{\alpha\beta}=\lambda_r+i\lambda_i$ for the isotropic case where $t_{\alpha}=t$ denotes the QD induced complex Majorana tunneling for $\alpha=1,2,...M$.

Notably, for different $M_{\mathrm{tot}}$ and $M$, Eq.\eqref{eq2} can be further mapped to multichannel Kondo models with exact channel symmetry \cite{li_multichannel_2023,zazunov_transport_2014,altland_bethe_2014}. For the minimal $M_{\mathrm{tot}}=4$ case \cite{beri_topological_2012} with $t_{\alpha}=t$, Eq.\eqref{eq2} is exactly reduced to Eq.\eqref{eq3} with isotropic $J^{\prime}$, where $c_{k,\alpha}$ denotes the ``rotated" electron operator of the leads. For example, $J^{\prime}=2(\lambda_r+i\lambda_i)$ and $n=4$ are obtained for $M=3$, while $J^{\prime}=4(\lambda_r+i\lambda_i)$ and $n=2$ are found for $M=4$. In addition, when $t_{\alpha}$ has dependence on $\alpha$, anisotropic models with different $J^{\prime}_l$ can also be realized. In particular, if one tunes the Majorana tunneling coefficients to $t_1=t^{\star}_2=te^{i\theta}$, $t_3=t$, with $\theta$ being an angle that parametrizes the non-Hermiticity,  an anisotropic nHMCK model is realized with $J^{\prime}_1=J^{\prime\star}_2=J^{\prime}e^{i\theta}$, $J^{\prime}_3=J^{\prime}$, which is \(\mathcal{PT}\) symmetric. Thus, the setup provides a controllable platform for exploring non-Hermitian Kondo physics under tunable symmetries. 

%Taking $M=3$ as an example, the lead electron operators join to form a spin-1 generator. Meanwhile, the Majorana modes encode an $\mathrm{SU}(2)$ Lie algebra, thus forming an effective pseudospin represented by spin-1/2 Pauli matrices.  Using the Kac-Moody algebra duality between $\mathrm{SO}(3)_1$ and $\mathrm{SU}(2)_4$  current~\cite{fabrizio_toulouse_1994}, this is mapped to a four-channel Kondo  model.
%For $M=4$, the model has an $\mathrm{SO}(4)$ symmetry which contains two sets of  $\mathrm{SU}(2)$, giving rise to an $\mathrm{SU}(2)_2$ two-channel Kondo problem. 
%Dualities generally exist for other $M$ ($M<M_{\mathrm{tot}}$), leading to multichannel $\mathrm{SU}(N)$ Kondo problems with exact channel symmetry \cite{li_multichannel_2023,zazunov_transport_2014,altland_bethe_2014}.

We now present a renormalization group (RG) analysis of Eq.\eqref{eq3}.  To the third-order perturbation, the RG flow equation is derived as $\frac{dg}{db}=g^2-ng^3/2$ , where $b$ is the RG scaling parameter, $g=\nu_0J^{\prime}=g_r+ig_i=ge^{i\theta}$, with $\nu_0$ being the density of states of the leads. As shown by Fig.\ref{fig3}(b), the flow diagram in the complex plane reveals a weak-coupling FP located at $(g_r,g_i)=(0,0)$ and a strong-coupling FP at $(g_r,g_i)=(2/n,0)$.  The strong-coupling FP describes the overscreened phase exhibiting nFL behavior \cite{nozieres1980kondo,cox_exotic_1998,patri_critical_2020}, while the weak-coupling FP depicts the local moment phase, where the effective pseudospin formed by the Majorana modes is decoupled from the conduction electrons. 

Around the weak-coupling FP, there emerges a circling RG flow, as originally identified by Ref.~\cite{nakagawa_non-hermitian_2018}. The system flows back to the weak-coupling FP even for antiferromagnetic interaction $g_r>0$. This reflects the fact that the dissipation favors decoupling rather than screening. As shown by the zoom-in data in Fig.\ref{fig3}(c), at high temperatures, the RG flow away from $(0,0)$ displays a universal growth, $g_r,g_i\sim \frac{1}{\mathrm{ln}(T/T_K)}$. Whereas, at low temperatures, a universal decreasing behavior, $g_r\sim g_i\sim \frac{1}{\mathrm{ln}(T_K/T)}$, is derived for the flow towards $(0,0)$. Note that the perturbative analysis well describes the low-energy physics, as $|g|\ll1$ is maintained in this weak-coupling regime.

%Furthermore, from the RG flow equation, a critical line (marked by yellow in Fig.\ref{fig3}(b)) separating the two FPs can be derived \cite{supp}, i.e., $\frac{\mathrm{sin}(\mathrm{arg}~g)}{|g|}=\frac{n}{2}\left[\pi-\mathrm{arg}~g+\mathrm{arg}(1-ng/2)\right]$ , where $\mathrm{arg}~g=\mathrm{arctan}(g_i/g_r)$.  To numerically verify the above perturbative RG results, we perform an exact diagonalization calculation of the nHMCK for $n=2$. The results in Fig.\ref{fig3}(e) clearly indicate the existence of critical points separating the two FPs.

Away from the weak-coupling FP, the situation is more subtle,  as the perturbative analysis could breakdown. Along the red arrow in Fig.\ref{fig3}(b), a critical curve (marked by yellow) occurs, separating the weak and strong non-Hermiticity region. Notably, for an intermediate non-Hermiticity region, Bethe ansatz studies on the single-channel case revealed an interesting Yu-Shiba-Rusinov (YSR)-like phase \cite{PhysRevB.111.L201106}.  Here, we solve the Bethe ansatz equation for the multichannel model without \(\mathcal{PT}\) symmetry, and find that the YSR-like phase still exists in this region (Sec.V of Ref.~\cite{Article}.). In the weak non-Hermiticity region with small $\theta$, although the perturbative RG predicts the flow towards $(g_r,g_i)=(2/n,0)$, non-perturbative calculations are indispensable to fully understand the low-energy Kondo physics. Thus, we further utilize the non-Hermitian NRG~\cite{19td-1k9s} to solve Eq.\eqref{eq3}, taking the $n=2$ case as an example. The details of our NRG method are presented in Ref.~\cite{Article}.

As shown by Fig.\ref{fig3}(e)(f), the real spectrum at large RG step coincides with that of the Hermitian case for small $\theta$, while significant deviations are observed for large $\theta$. Meanwhile, the impurity entropy saturates to $\mathrm{ln}2/2$ and $\mathrm{ln}2$ for small and large $\theta$, respectively (Fig.\ref{fig3}(d)). These indicate that the system indeed flows to the overscreened (unscreened) phase under weak (strong) non-Hermiticity. For small $\theta$, the entropy decreases monotonously with temperature (blue curve in Fig.\ref{fig3}(d)), displaying the standard Kondo regime behavior. For large $\theta$, the local moment phase is identified for both the high and low temperature regime, again validating the circling RG flow in Fig.\ref{fig3}(b) ~\cite{nakagawa_non-hermitian_2018}. Remarkably,  for both large and intermediate $\theta$, the impurity entropy exhibits a non-monotonous evolution and can exceed the free local moment entropy, $\ln2$, in an intermediate temperature range (red curve in Fig.\ref{fig3}(d)). Such anomalous behaviors were also reported in superconductors with an impurity ~\cite{kattel2025multichannelkondoeffectsuperconducting,kattel2025thermodynamicssplithilbertspace} . They reflect the fact that additional degrees of freedom emerge and contribute to the impurity entropy, providing clear evidence of the YSR phase predicted by Ref.~\cite{PhysRevB.111.L201106}. 

%We further note that the imaginary spectrum in Fig.\ref{fig3}(e)(f) generally remains finite. This implies that the nature of the strong-coupling fixed point is beyond the Hermitian Kondo framework, which could be described by complex CFTs~\cite{tang_reclaiming_2024,han_complex_2023} that require further studies. In the following, to obtain reliable experimental signatures of the FPs, we will focus on the Kondo conductance by imposing \(\mathcal{PT}\) symmetry on Eq.\eqref{eq3}.
 
{\color{blue}\emph{Non-Hermiticity induced anomalous Kondo conductance.}}-- To obtain reliable transport properties associated with the FPSs, we will now focus on the nHMCK model with \(\mathcal{PT}\) symmetry. We solve the \(\mathcal{PT}\) symmetric  $n=2$  model,  using non-Hermitian NRG. As shown by Fig.\ref{fig4}(a), for small $\theta$, the real part of the energy spectrum is similar with that without \(\mathcal{PT}\) symmetry (Fig.\ref{fig3}(e)). Whereas, for large $\theta$, the deviation of the real spectrum from the Hermitian case is completely suppressed at large RG step (Fig.\ref{fig4}(b)). Moreover, the imaginary spectrum also vanishes in low-energy (or equivalently, large step), in sharp contrast to Fig.\ref{fig3}(e)(f). Since the imaginary spectrum is negligible in the \(\mathcal{PT}\) unbroken region, the strong-coupling phase predicted above is validated. Furthermore, the energy levels and the degeneracy obtained by non-Hermitian NRG exactly coincides with those predicted by BCFT, as shown by Fig.\ref{fig4}(b). This clearly justifies the applicability of BCFT for the strong-coupling phase, as long as \(\mathcal{PT}\) symmetry is preserved.
% More numerical results of thermodynamic quantities are presented in Sec.IV of Supplemental Materials

We now investigate the transport properties \cite{PhysRevB.110.045138,PRXQuantum.3.030308,PhysRevLett.134.156601} associated with both phases. For the weak-coupling phase, a perturbative non-Hermitian Kubo formula works well  \cite{sticlet_kubo_2022}, while for the strong-coupling phase, the non-perturbative BCFT is applicable, as demonstrated above \cite{affleck_current_1990,affleck_exact_1993,affleck_critical_1991,affleck_kondo_1991}. We first apply the linear response theory to study Eq.\eqref{eq2}. The Hamiltonian can be formally written as $H=H_r-iH_i$, where $H_r$ ($H_i$) denotes the Hermitian (anti-Hermitian) part. The evolution of the density matrix is given by $\rho(t)=e^{-iHt}\rho(0)e^{iH^{\dagger}t}/\mathrm{Tr}(e^{-iHt}\rho(0)e^{iH^{\dagger}t})$, where the denominator ensures the proper normalization \cite{brody_mixed-state_2012}. Using the non-Hermitian Kubo formula \cite{sticlet_kubo_2022}, the conductance (between the $\alpha$-th and $\beta$-th lead, $\alpha\neq\beta$) is obtained as, 
\begin{equation}\label{eqKubo}
G_{\alpha\beta}=-\mathrm{Re}\lim_{\omega\rightarrow0}\frac{1}{\omega}\int^{\infty}_0dte^{i\omega_+t}\langle[\tilde{I}_{\alpha}(t),\tilde{I}_{\beta}(0)]\rangle_{\sim},
\end{equation}
and the current-current correlation is
\begin{equation}\label{eqcurrentcorrelation}
\begin{split}
    \langle[\tilde{I}_{\alpha}(t),\tilde{I}_{\beta}(0)]&\rangle_{\sim}=\mathrm{Tr}
\{\rho(0)[\tilde{I}_{\alpha}(t),\tilde{I}_{\beta}(0)]_{\sim}\\
   & -\langle \tilde{I}_{\alpha}(0)\rangle_0[e^{iH^{\dagger}t}e^{iHt},\tilde{I}_{\beta}(0)]\rho(0)\}/\mathrm{Tr\rho(0)},
    \end{split}
\end{equation}
where the first term in Eq.\eqref{eqcurrentcorrelation} is the generalization of the Kubo formula for the non-Hermitian setting, with  $[A,B]_{\sim}=AB-B^{\dagger}A$. The second term arises due to the nonunitary dynamics, and  $\langle \tilde{I}_{\alpha}(0)\rangle_0=\mathrm{Tr}(\tilde{I}_{\alpha}(0)\rho(0))/\mathrm{Tr}\rho(0)$. Hereby, we have assumed the quasi-unitary state evolution around the FPs, i.e., $\rho(t)\sim\rho(0)/\mathrm{Tr}(\rho(0))$. This is justified because $\lambda_i$ is small around the weak-coupling FP.  
\begin{figure}
\includegraphics[width=\linewidth]{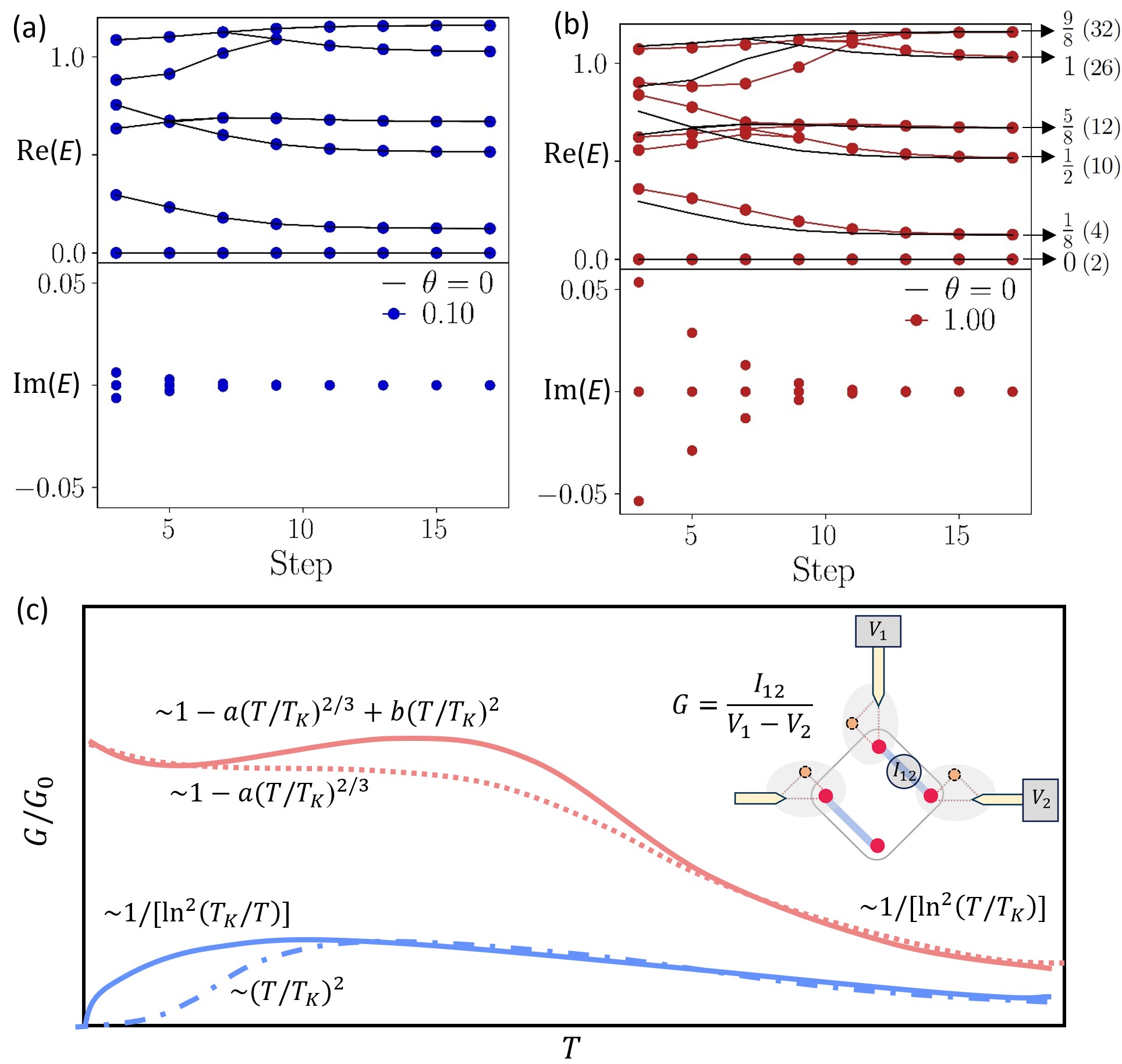}
\caption{\label{fig4}(a) and (b) show the NRG results of the real and imaginary energy spectrum for the \(\mathcal{PT}\) symmetric nHMCK model.  The spectrum is shown for the weak (a) and strong (b) non-Hermiticity case, respectively. The obtained energy levels and degeneracies exactly coincide with those predicted by BCFT (highlighted by arrows in (b)).   (c) The conductance of the nHMCK ($n=4$). The red and blue solid curves correspond to the strong- and weak-coupling phases, respectively. The red dashed curve shows the strong-coupling conductance for the Hermitian 4-channel Kondo model. The blue dashed-dot curve is the conductance in the unscreened channel of an anisotropic multichannel Kondo model \cite{PhysRevB.83.245308}, which is shown for the purpose of comparison. The inset depicts the measurement setup.}
\end{figure}

The current operator is obtained via $\tilde{I}_{\alpha}=-\frac{dN_{\alpha}}{dt}=-i[H_r,N_{\alpha}]-[H_i,N_{\alpha}]_+=I_{\alpha,r}+I_{\alpha,i}$. The specific forms of $I_{\alpha,r}$ and $I_{\alpha,i}$  are included in Ref.~\cite{Article}. Inserting $\tilde{I}_{\alpha}$ into Eq.\eqref{eqcurrentcorrelation}, we find that the conductance consists of three terms  around the FPs, $G_{\alpha\beta}=G_{0,\alpha\beta}+G_{\star,\alpha\beta}+G_{\langle\rangle,\alpha\beta}$, respectively arising from the correlation functions,  $C_r(t,0)\sim\langle [I_{\alpha,r}(t),I_{\beta,r}(0)]\rangle_0$, $C_i(t,0)\sim\langle [I_{\alpha,i}(t),I_{\beta,i}(0)]\rangle_0$, and $C_{\langle\rangle}(t,0)\sim\langle[H_i,I_{\beta,i}(0)]\rangle$. 

At the weak-coupling FP, we obtain $C_r(t,0)\sim\lambda^2_r\sim g^2_r$, and $C_i(t,0),C_{\langle\rangle}(t,0)\sim \lambda^2_i\sim g^2_i$. Recalling the scaling behavior of $g_r$ and $g_i$ around the weak-coupling FP, it can be derived that $G_{\alpha\beta}(T)\sim\frac{1}{\mathrm{ln}^2(T/T_K)}$ at high temperatures and $G_{\alpha\beta}(T)\sim\frac{1}{\mathrm{ln}^2(T_K/T)}$ at low temperatures. The conductance is plotted by the blue solid curve in Fig.\ref{fig4}(c). As shown, the conductance flows to zero and meanwhile obeys $\sim\frac{1}{\mathrm{ln}^2(T_K/T)}$. Remarkably, this is distinct from all previously reported Kondo systems. Although a similar temperature dependence was found for underscreened Kondo models \cite{posazhennikova_anomalous_2005,coleman_singular_2003,mehta_regular_2005}, the conductance therein saturates to finite values for $T\rightarrow0$. In addition, although zero conductance for $T\rightarrow0$ occurs in the unscreened channel of anisotropic multichannel Kondo models \cite{PhysRevB.83.245308,andrei_fermi-_1995}, Fermi-liquid-type behavior, $\sim(T/T_K)^2$, emerges at low temperatures in that case, as indicated by the blue dashed-dot curve in Fig.\ref{fig4}(c). Thus, the conductance here is unique, as it is attributed to a completely different mechanism--the nH-induced decoupling. 

For the strong-coupling phase, the correlation functions above can be evaluated using BCFT.
%~\cite{supp}
In CFT, the Kondo model in Eq.\eqref{eq3} is mapped to an $\mathrm{SU}(2)_n$ Wess-Zumino-Witten (WZW) model with conformal invariant boundaries \cite{affleck_current_1990,affleck_kondo_1991}. The WZW model is characterized by the spin of the highest weight state, i.e., $j=0,1/2,...,n/2$, whose descendants constitute the conformal towers labeled by $j$. The spin-$j$ primary field, $\phi_j$, has the scaling dimension, $\Delta_j=\frac{j(j+1)}{n+2}$.  In the following, we will focus on the minimal $M=3$ case, which exhibits the $\mathrm{SU}(2)_4$ algebra. 

The low-energy thermodynamics is determined by the leading irrelevant operators (LIOs), which should satisfy the Kondo boundary condition and meanwhile respect the symmetries of Eq.\eqref{eq3}, including the \(\mathcal{PT}\) symmetry. These operators are generated by the $\mathrm{SU}(2)_4$ double fusion with the spin-1/2 primary \cite{affleck_kondo_1991}, which sends the $j=0$ primary to $j=0,1$ and $j=2$ primary to $j=1,2$. In the Hermitian case, the LIO is constructed as the first descendant of the $j=1$ primary, $\mathbf{J}_{-1}\phi_{j=1}$, which has scaling dimension $\Delta=\frac{4}{3}$.  Further considering that the neutrality condition on the vertex operators \cite{beri_topological_2012},  the LIO leads to a low-energy conductance exhibiting non-Fermi liquid behavior, i.e.,  $G_{\alpha\beta(T)}=G_0[1-a T^{2(\Delta-1)}]=G_0(1-a T^{2/3})$ \cite{beri_topological_2012,zazunov_transport_2014}, where $G_0=2e^2/3h$ and $a$ is a nonuniversal coefficient.

Interestingly, for the non-Hermitian model here, there emerges a new \(\mathcal{PT}\)-preserving irrelevant operator,  i.e., $i\mathbf{J}_{-1}\phi_{j=2}$, which has the scaling dimension $\Delta=2$. Considering the second-order contribution of this LIO, further correction is generated for the conductance, leading to $G_{\alpha\beta}(T)=G_0(1-aT^{2/3}+bT^2)$, where $b$ is another non-universal coefficient \footnote{Note that the different sign in front of $b$ and $a$ comes from the square of $i$. }, arising from the non-Hermitian effect.  The non-Hermiticity introduces a new irrelevant operator around the non-Fermi liquid FP, which plays a crucial role in  reshaping the Kondo conductance. 
The conductance in the strong-coupling regime is shown by the red solid curve in Fig.\ref{fig4}(c). As shown, in addition to the non-Fermi liquid feature,  an anomalous upturn could generally emerge with increasing $T$. Such a behavior is again absent in all previously studied Kondo systems \cite{nozieres1980kondo,hewson_kondo_1997,nakagawa_non-hermitian_2018}, serving as a characteristic feature driven by the intertwined effects of nH and nFL.

%We highlight that the scalings are essentially different from conventional Kondo problems \cite{nozieres1980kondo,hewson_kondo_1997} because the FPs are enriched by non-Hermiticity. Moreover, they are also distinct from the non-Hermitian Kondo effect \cite{nakagawa_non-hermitian_2018} due to the intrinsic interplay between nH and nFL around the strong-coupling FP. 

{\color{blue}\emph{Conclusion and discussion.}}-- 
%Therefore, the  anomalous upturn (red curve in Fig.4) directly reflects the exotic interplay between non-Fermi liquid and non-Hermiticity. In addition, as discussed above, although the non-Fermi liquid behaviors are absent in the weak-coupling regime, the non-Hermiticity can still enrich the Kondo conductance in a nontrivial way, as exemplified by the blue curve in Fig.4.Anomalous transport associated with Kondo effect is a topical field that has been advancing condensed matter physics for decades. Here,
We have shown that non-Hermiticity can generate unconventional Kondo conductance behaviors. Particularly, when nFL meets nH, an anomalous upturn would occur, originating from an emergent BCFT with non-Hermitian boundary operators. The non-Hermitian LIO identified here is unique for Kondo systems that simultaneously display nH and nFL. For example, for the non-Hermitian Fermi liquid FP (which occurs for $n=1$) , the impurity is exactly screened, and the LIOs around the strong-coupling FP are simply four-fermion operators without imaginary couplings \footnote{The four-fermion operators with imaginary couplings break \(\mathcal{PT}\), thus are prohibited around the FP.}. Here, we emphasize that these BCFT results are applicable only to \(\mathcal{PT}\) symmetric model, while for the asymmetric cases, our NRG calculations have confirmed the emerging YSR phase \cite{PhysRevB.111.L201106,PhysRevB.111.224407}, beyond BCFT descriptions. In addition, it is worthwhile to consider generalizations to other correlated systems exhibiting nFL properties, such as the Luttinger liquid \cite{lee_kondo_1992,frojdh_kondo_1995,furusaki_kondo_1994,han_complex_2023,tang_reclaiming_2024}.  Furthermore, we remark that multi-orbital cold-atom systems also provide a feasible platform to realize our predicted Kondo phenomena, as they can be engineered to host both localized and itinerant degrees of freedom ~\cite{Gorshkov_2010,PhysRevLett.120.143601}. Designable dissipation and system-integrable monitoring equipments are also achievable in this platform ~\cite{doi:10.1126/science.1155309,10.1093/ptep/ptaa094,Bouganne_2020,skin_2025,universal_2025,Kondozeno_2025}. Last, our predicted anomalous conductance could serve as an additional signature of Majorana fermions. Given its strong-correlation and nonlocal origin, this signature may help distinguish Majorana modes from trivial bound states in experimental measurements.

%the Majorana-based platform--experimentally accessible in practice--guarantees exact channel symmetry and thus the predicted anomalous Kondo conductance. The latter, therefore, would offer, serving as both a probe for their existence and a guide for their potential applications.

\begin{acknowledgments}
W. Z. Y and Y. C contributed equally to this work. We acknowledge Wei-Qiang Chen for inspiring discussions. This work was supported by the National Natural Science Foundation of China (No.12322402, No.12274206, No.125B2076), the Scientific Research Innovation Capability Support Project for Young Faculty (SRICSPYF-ZY2025164), the National R\&D Program of China (2024YFA1410500, 2022YFA1403601),  the Quantum Science and Technology-National Science and Technology Major Project (Grant No.2021ZD0302800), the Natural Science Foundation of Jiangsu Province (No.BK20233001), and the Fundamental Research Funds for the Central Universities (KG202501).
\end{acknowledgments}

\bibliographystyle{apsrev4-1-etal-title_10authors}
%\bibliography{KONDO.bib}

\end{document}